\documentclass[12pt,14paper]{article}
\usepackage[reqno]{amsmath}
\usepackage{epsfig}
\usepackage{array}
\usepackage{float}
\usepackage{lscape,graphicx} 
\usepackage{graphics}
\usepackage{amssymb}
\usepackage{amsmath}
\usepackage{slashed}
\usepackage{color}
\usepackage{url}
\usepackage{multirow}
\usepackage{advdate}
\usepackage{datenumber}
\usepackage{ulem}

\newcommand{\GeV}{\mbox{$~{\rm GeV}$}}
\newcommand{\TeV}{\mbox{$~{\rm TeV}$}}

\newcommand{\bea}{\begin{equation}\begin{array}{c}}
\newcommand{\eea}{\end{array}\end{equation}}
\newcommand{\ea}{\end{array}}

\newcommand{\beq}{\begin{equation}}
\newcommand{\eeq}{\end{equation}}
\newcommand{\bad}{\begin{array}{ccc}}

\newcommand{\ba}{\begin{array}{c}}

\begin{document}

\title{On the impact of the Higgs boson on the production \\ of exotic particles at the LHC}

\author{A.~G.~Hessler, A.~Ibarra, E.~Molinaro and S.~Vogl\\\\
{\it Physik-Department T30d, Technische Universit\"at M\"unchen,}\\ 
{\it{James-Franck-Stra{\ss}e, 85748 Garching, Germany.}}}

\date{}

\maketitle
\vspace*{-10cm}
\begin{flushright}
\texttt{\footnotesize TUM-HEP 950/14}\\[-1mm]
\texttt{\footnotesize FLAVOUR(267104)-ERC-68} 
\end{flushright}
\vspace*{8.5cm}
\thispagestyle{empty}
\begin{abstract}
Many new physics models contain new particles that interact with the Higgs boson.  These particles could be produced at the LHC via gluon-gluon fusion with an off-shell Higgs and, if charged under a gauge group, also via gluon-gluon fusion with an off-shell $Z$ boson and the  Drell-Yan process.  We consider in this paper simplified scenarios where the Standard Model is extended by one scalar or fermionic field that interacts with the Higgs boson and we evaluate the impact of the Higgs interaction on the production of the exotic particles at the LHC. This analysis applies in particular to TeV scale seesaw scenarios of neutrino mass generation.
\end{abstract}

\section{Introduction}
\label{sec:Introduction}

The discovery by the ATLAS and CMS collaborations of a new boson with a mass of approximately 125 GeV has opened a new era in Particle Physics~\cite{Aad:2012tfa,Chatrchyan:2012ufa}. On the one hand, the measured properties of the new bosonic particle are in remarkable agreement with those expected for the long-awaited Higgs boson~\cite{CMS:2014ega}, predicted by the mechanism of spontaneous breaking of the electroweak symmetry to generate the gauge boson, charged lepton and quark masses.  On the other hand, the discovery of the new boson also constitutes evidence for a new fundamental interaction in Nature. From the theoretical point of view, this new interaction plays a crucial role in preserving the perturbative unitarity of the theory at high energies and in guaranteeing the renormalizability of the Standard Model (SM). However, it also has many phenomenological implications, such as the imprint it leaves in the electroweak precision measurements which led to indirect hints for the existence of the Higgs boson already in the LEP era~\cite{Alcaraz:2007ri}, or its production at the LHC and decay into other particles, that led to its identification. 

New particles beyond the SM could also couple to the Higgs field. More specifically, the gauge and the Lorentz symmetries allow interaction terms of the Higgs doublet $H$ with new scalars $S_i$ of the form $|H|^2 |S_i|^2$, among other terms, or with fermions $\Psi_1$, $\Psi_2$ of the form $\bar \Psi_1 \Psi_2 \,H$. 
We consider for simplicity a minimal number of new degrees of freedom in addition to the SM, that is we explicitly analyse the production in proton-proton colliders
of new particles in models with just one extra $SU(2)_L$ scalar multiplet or one new chiral fermion.
Under these assumptions, it can be checked that the scalar Higgs portal term is present in the potential for any assignment of the gauge quantum numbers for the scalar field, while  the fermionic Higgs portal term exists only if either $\Psi_1$ or $\Psi_2$ is one SM lepton and 
the new chiral fermion has gauge quantum numbers under $SU(3)_c\times SU(2)_L\times U(1)_Y$ equal to $(1, 1,0)$ or $(1,3,0)$ (these two cases correspond to the renown type I \cite{typeI} and type III \cite{typeIII} seesaw mechanisms, respectively). Thus new scalars  or fermions can be produced at the LHC via their electroweak gauge interactions with the partons inside the protons  (see, {\it e.g.}, \cite{delAguila:1990yw,delAguila:2008cj,Franceschini:2008pz}). However, this is not the only possibility for production, since the newly discovered Higgs particle can also mediate the production of exotic scalars or fermions. Due to the  fairly good recent determination of the mass and some of the couplings of the Higgs boson, a quantitative analysis of the impact of the Higgs boson on the production rate of exotic particles at the LHC has  now become possible. 
Therefore, we perform in this work a quantitative comparison  between Higgs-mediated channels and the well-known electroweak production channels. 
To the best of our knowledge, such a systematic analysis has not been done before. 

The paper is organized as follows: In section \ref{Sec:Basics} we recapitulate basic elements of the formalism to calculate the production cross-section of exotic particles at the Large Hadron Collider. In section \ref{Sec:scalar} we study the production of exotic scalars, while in section \ref{Sec:fermion} the production of exotic fermions is analyzed, considering in both sections a center-of-mass energy of 8 and 14 TeV. In section \ref{Sec:100TeV} we briefly comment on the prospects to observe exotic particles in a hypothetical proton-proton collider operating at a center-of-mass energy of 100 TeV and, lastly, in section \ref{Sec:conclusions} we present our conclusions.

\section{Heavy particle production at the LHC}
\label{Sec:Basics}

New physics states interacting with the Higgs boson can be produced at the LHC via the gluon-gluon fusion (ggF) vertex with a Higgs particle.\footnote{We will not consider here the production via vector boson fusion, although this process could give a non-negligible contribution to the cross-section at large momentum transfer~\cite{Djouadi:2005gi}.} The leading order contribution to this vertex consists of a triangular diagram with the top quark in the loop and subleading corrections due to the  bottom quark, while lighter flavours are essentially negligible. We follow the common procedure in the literature~\cite{Spira:1995rr} and take the gluon-gluon-Higgs (ggH) process into account by introducing an effective Lagrangian for the Higgs interaction with gluons
\begin{eqnarray}
\mathcal{L}_{\mathrm{H,eff}}= \frac{1}{4}\, G_H\, G_{\mu \nu}^a\, G^{a\, \mu \nu}\, \frac{h}{v},
\end{eqnarray}  
where $G_H$ is an effective coupling and $v\simeq 246$ GeV is the Higgs vacuum expectation value (vev),  while $G_{\mu \nu}^a$ and $h$ correspond to the gluon field strength tensor and the Higgs field respectively. 
 
The effective coupling can be determined by matching the amplitude of the effective theory to the  corresponding amplitude in the SM and is given, keeping only the contribution of the top quark, by
\begin{equation}
	G_{H} \;=\; \frac{\alpha_{S}}{2\,\pi }\,\left|F_{H}\left(\frac{4\,m^{2}_{t}}{P^2}\right)\right|\,, \label{GHeff}
\end{equation}
where  $\alpha_{S}$ is the strong coupling constant and $F_{H}$ corresponds to the well-known triangle form factor~\cite{Plehn:1996wb} which depends on the top quark mass $m_{t} = 173.5 \GeV $ and the momentum scale of the process $P$. Depending on whether the process considered is on-shell Higgs production or new physics production from a Higgs boson at arbitrary virtuality,  the scale $P$ is either set by  the  Higgs mass $m_{h}\simeq 125$ GeV or  by the total partonic center-of-mass (c.o.m.) energy  $\sqrt{\hat{s}}$. The form factor can be calculated analytically and is given by 
\begin{eqnarray}\label{loopfunction}
	F_{H}(\tau) & = & \tau\,\left[1\,+\,(1-\tau)\,f(\tau)\right]\,,
\end{eqnarray}
with
\begin{equation}
	f(\tau)\;=\begin{cases}
	\displaystyle{\arcsin^{2}\frac{1}{\sqrt{\tau}}}& {\rm if}~\tau\geq 1\\
	\displaystyle{-\frac{1}{4}\,\left[\log\left(\frac{1\,+\,\sqrt{1\,-\,\tau}}{1\,-\,\sqrt{1\,-\,\tau}}\right)\,-\,i\,\pi\right]^{2}}&{\rm if}~\tau< 1\,.
        \end{cases}
\end{equation}

Furthermore, an additional contribution to gluon-gluon fusion can arise from the gluon-gluon-$Z$ (ggZ) triangle diagram\footnote{Furry's theorem forbids the analogous process with an intermediate photon.}. Since we are only interested in the production from real on-shell gluons
in the initial state, the general form of the ggZ vertex originally derived in \cite{Bell:1969ts}  can be simplified and the effective Lagrangian reads
\begin{eqnarray}
\mathcal{L}_{\mathrm{Z,eff}} =  \frac{1}{4}\,G_Z \,\partial_\alpha\,Z^\alpha\,G^a_{\mu \nu}\,\tilde{G}^{a\mu\nu}\,,
\end{eqnarray} 
where $\tilde{G}^{\mu \nu}$ is the dual of the field strength tensor. The effective coupling $G_Z$ is given by~\cite{Dicus:1991wj}
\begin{equation}
G_Z = \frac{\alpha_s}{2\,\pi}\,\frac{e}{s_w\,c_w}\,\left|\sum_Q (-1)_Q\,F_Z(\hat{s},m_Q^2)\right|\,,
\label{eq:GZ}
\end{equation}
where  $s_w$ and $c_w$ correspond to the sine and the cosine of the Weinberg angle, while the form factor reads
\begin{equation}\label{ggZloopfactor}
F_Z(\hat{s},m_Q^2)=\frac{1}{\hat{s}}\left\{1+\frac{2\,m_Q^2}{\hat{s}}\int_0^1\frac{dx}{x}\log\left[1-\frac{\hat{s}}{m_Q^2}\,x\,(1-x)\right]\right\}\,.
\end{equation}
The sum runs over all quark flavours $Q$ present in the triangle loop and $(-1)_Q$ equals $+1$ for up-type quarks and $-1$ for down-type quarks. The first term in the from factor in Eq.~\eqref{ggZloopfactor} is flavour-independent and therefore drops out once the sum over complete generations of quarks is performed. As the second term is proportional to the square of the quark mass, the top contribution will be once again dominant and lighter quark flavours can be safely neglected.

With the effective couplings $G_{H}$ and $G_Z$, the partonic cross-sections can be calculated in different models; the corresponding expressions for the production of exotic scalar and fermionic particles will be reported in sections \ref{Sec:scalar} and \ref{Sec:fermion}. We point out that the Higgs- and $Z$-mediated gluon-gluon fusion amplitudes do not interfere due to their different Lorentz index structure.

At the LHC, the production cross-section of the final  state $X_1 X_2$, where $X_1$ and $X_2$ are not necessarily different particles,  is the result of the convolution of the partonic cross-section  with the corresponding parton distribution functions (PDFs) {(see {\it e.g.} \cite{Djouadi:2005gi})}. More specifically for the case of gluon-gluon fusion,
\begin{equation}
	\sigma\left(p\,p\to X_1 X_2;\,s\right)=\int_{\tau_{s}}^1 \frac{d\mathcal{L}^{gg}}{d\tau }\,\sigma\left(g g\to X_1 X_2;\,\hat s=\tau s\right)\, d\tau\,,\label{DYCS}
\end{equation}
where $\tau_{s}\equiv (m_{X_1} + m_{X_2})^2/s$, $\sqrt{s}$ is the center-of-mass energy of the proton-proton collision and ${\cal L}^{gg}$ is the gluon luminosity function, defined as
\begin{equation}
    \frac{d\mathcal{L}^{gg}}{d\tau }=\int_{\tau }^1 \frac{g\left(x,\mu _F\right) g\left(\frac{\tau }{x},\mu_F\right)}{x} \, dx\,.
\label{lum1}
\end{equation}
The gluon PDF, $g(x,\mu_{F})$, depends on the fraction $x$ of the proton momentum carried by the parton and on the factorization scale $\mu_{F}$, which represents the scale at which the matching between the perturbative calculation of the matrix elements and the non-perturbative part related to the parton distribution functions is performed. In our analysis we will focus on the leading order contributions to the production 
cross-section of exotic particles via  gluon-gluon fusion, however, it should be borne in mind that higher order corrections to this production channel can significantly enhance the cross-section. For example, comparing  leading order results to state-of-the-art calculations for single Higgs production \cite{xsecGroup}, one finds $ K \equiv \sigma_{\rm NNLO+NNLL} / \sigma_{\rm LO} \sim 3$ at 14 TeV.  

The same final state can be generated via the Drell-Yan (DY) process, $i.e.$ by  the exchange of an off-shell photon and/or $Z$ boson in the $s$-channel in a quark-antiquark collision. The corresponding production cross-section via this channel reads
\begin{equation}
	\sigma\left(p\,p\to X_1 X_2;\,s\right)=\sum_{q=u,d}\int_{\tau_{s}}^1 \frac{d\mathcal{L}^{q\bar{q}}}{d\tau }\,\sigma\left(q\bar{q}\to X_1 X_2;\hat s=\tau s\right)\, d\tau\,,
	\label{DYCS2}
\end{equation}
where the quark luminosity function is
\begin{equation}
 \frac{d\mathcal{L}^{q\bar{q}}}{d\tau }=\int_{\tau }^1 \frac{q\left(x,\mu _F\right) \bar{q}\left(\frac{\tau }{x},\mu_F\right)+
 		q\left(\frac{\tau }{x},\mu _F\right) \bar{q}\left(x,\mu_F\right)}{x} \, dx\,, \label{lum2}
\end{equation}
with $q(x,\mu_{F})$  and  $\bar q(x,\mu_{F})$ the quark and antiquark PDFs, respectively.  The relative importance of the two production mechanisms depends on the details of the new physics model. In the following, we discuss separately the case of  scalar particle production and of  fermionic particle production.

\section{Scalar multiplet production at the LHC}
\label{Sec:scalar}

We consider a minimal extension of the SM with one additional  complex scalar field $S$, assumed to be uncoloured, and which is a generic multiplet of $SU(2)_{L}$ with weak isospin $T$ and hypercharge $Y$ (we will not consider new particles which are charged under $SU(3)_c$ as their production is completely dominated by strong interactions). The scalar field $S$ has $n=2\,T+1$ components, which  are
\begin{equation}
	S=\left(S_{1},\ldots ,S_{n}\right)^{T}\,.
\end{equation} 
Then, all the interactions of $S$ with the SM electroweak gauge bosons and the Higgs doublet $H$ are given by the Lagrangian  
\begin{eqnarray}
\mathcal{L}_{S} &=& \left(D_{\mu}\,S\right)^{\dagger}\,\left(D^{\mu}\,S\right)\,-\,V(S,H)\,.\label{LS}
\end{eqnarray}
In the charge basis the covariant derivative $D_{\mu}$ takes the usual form,
\begin{equation}
	D_{\mu}\;=\; \partial_{\mu}\,-\,i\frac{g}{\sqrt{2}}\,\left(W_{\mu}^{+}T^{+}\,+\,W_{\mu}^{-}T^{-}\right)-
	i\frac{1}{\sqrt{g^{2}+g'^{2}}}\,Z_{\mu}\left(g^{2}T^{3}-g'^{2}\frac{Y}{2}\right)
	-\,i\frac{g g'}{\sqrt{g^{2}+g'^{2}}}\,A_{\mu}\,Q\,,\nonumber
\end{equation}
where we define the electric charge $Q=T^{3}+Y/2$.

We consider as  benchmark scenario in our discussion the most generic renormalizable scalar potential 
$V(S,H)$ \cite{Hambye:2009pw} invariant under a global $U(1)$ symmetry:\footnote{In the case of a real multiplet, the kinetic term in Eq.~(\ref{LS}) is multiplied by a factor 1/2 and the operator $S^{\dagger}T^{a}S$ in Eq.~(\ref{VS}) is identically zero, so that the components of a real multiplet are degenerate in mass.}
\begin{eqnarray}
	V(S,H) & = & \mu_H^2\,H^\dag H+f_1\left(H^{\dagger}H\right)^{2}+\mu_{S}^{2}\,S^{\dagger}S+f_{2}\left(S^{\dagger}S\right)^{2}+f_{3}\left(H^{\dagger}H\right)\left(S^{\dagger}S\right)\nonumber\\
			&&+f_{4}\left(H^{\dagger}\tau^{a}H\right)\left(S^{\dagger}T^{a}S\right)+f_{5}\left(S^{\dagger}T^{a}S\right)\left(S^{\dagger}T^{a}S\right)\,.\label{VS}
\end{eqnarray}
We define  $T^{a}$ ($\tau^{a}$) for $a=1,2,3$ as the generators of $SU(2)_{L}$ in a generic (the fundamental)  representation, such that, in a basis in which $T^3$ is diagonal, $T^{3} S_{1} = T S_{1}$ and $T^{3} S_{n}=-T S_{n}$.

After electroweak symmetry breaking, one of the components of $S$ is neutral for particular choices of the weak isospin and the hypercharge. In the following, we assume that the neutral component of the scalar multiplet does not acquire a vev (or acquires only a very small vev,  as in the examples that will be shown later), such that the formalism for the gluon-gluon fusion production mechanism described in section \ref{Sec:Basics} can be applied.

The coupling $f_{4}$ in Eq.~(\ref{VS}) controls the mass splitting of the multiplet components, at tree level.\footnote{If the global $U(1)$ is explicitly broken
to an accidental $Z_{2}$ symmetry, then for $n$ even and $Y=1$ the operator $(H^{\dagger}\tau^{a}\widetilde{H})(S^{\dagger}T^{a}\widetilde{S})$
is allowed, where $\widetilde{H}$ and $\widetilde{S}$ are the conjugate scalar multiplets. This operator provides an independent contribution to the relative mass splitting of the $S$ components.} In the following, we will classify the multiplet components by the electric charge, where we refer to the component with charge equal to $Q$ by $S^{(Q)}$. In this notation, the tree level squared mass of each eigenstate after electroweak symmetry breaking is given by
\begin{eqnarray}
	m_{S^{(Q)}}^{2} &=& \mu_{S}^{2}\,-\,\frac{1}{4}\,\Lambda_{Q}\,v^{2}\,,\label{eq:mQ}
\end{eqnarray}
where the coupling $\Lambda_{Q}$ is defined as
\begin{eqnarray}
	\Lambda_{Q}&=&-\,2\,f_{3}\,+\,f_{4}\,T^3\;=\;-\,2\,f_{3}\,+\,f_{4}\,\left(Q-\frac{Y}{2}\right)\,,\label{LQ}
\end{eqnarray}
and controls the strength of the interaction of the Higgs boson with two scalars, namely
\begin{eqnarray}
	\mathcal{L}_{\rm int} &=& \frac{1}{2}\,\Lambda_{Q}\,v\,h\,S^{(Q)}\bar{S}^{(Q)} \,+\,\frac{1}{4}\,\Lambda_{Q}\,h^{2}\,S^{(Q)}\bar{S}^{(Q)}\,.
\end{eqnarray}
 Note that the mass splitting among the members of the multiplet, labelled by their electric charges, can be positive or negative depending on the sign of the quartic coupling~$f_4$.

The parameter space spanned by the multiplet masses and couplings is  constrained by various theoretical and phenomenological considerations, as we will briefly discuss now. Under the common requirement of vacuum stability, the following tree level conditions for the couplings must be fulfilled \cite{Hambye:2009pw}:
\begin{eqnarray}\label{LambdaQVacStabCond}
	f_{1,2} &>& 0\,,\label{LambdaQVacStabCond1}\\
	\Lambda_{Q} & < & 4\sqrt{f_1f_2} \; = \; 2\;\frac{m_{h}}{v}\;\sqrt{2\, f_{2}}\;\simeq\; 5\,,\label{LambdaQVacStabCond2}
\end{eqnarray}
where in the last equation the numerical value corresponds to the perturbativity limit $f_{2}=4\pi$. 

Unitarity bounds on scalar-scalar and scalar-gauge boson scattering amplitudes constrain the couplings in the scalar potential. These constraints are model dependent and must be derived for each multiplet dimension and choice of hypercharge and $U(1)$ or $Z_2$ charge from the contributing process amplitudes involving the corresponding couplings.
Constraints from tree level partial wave unitarity impose that $n\leq 8$ ($n\leq 9$) in the case of a complex (real) scalar multiplet (see, $e.g.$,  \cite{Hally:2012pu}). Furthermore, electroweak precision constraints, which are usually formulated in terms of the so-called oblique parameters $S$, $T$ and $U$ \cite{STU}, need to be taken into account. The most stringent constraint stems from the $T$ parameter, while $S$ and $U$ provide weaker limits.
Expressions for these in the case of an additional scalar multiplet of arbitrary size, which does not acquire a vacuum expectation value, have been calculated in \cite{Lavoura:1994}.
Also, additional charged scalars coupled to the Higgs contribute to the  $h \rightarrow \gamma \gamma$ rate \cite{Gunion:1989we}.
Concerning the mass spectrum,  the invisible width of the $Z$ boson typically implies a lower bound, $m_{S^{(Q)}}\gtrsim m_{Z}/2\simeq 45$ GeV.

Notice that a complex scalar multiplet can provide a viable dark matter candidate if it has a vanishing hypercharge --as dictated by direct detection constraints-- and if its lightest component is neutral. Interestingly, in the case of a scalar multiplet of dimension $n\ge7$ and $Y=0$, the neutral component is naturally long lived, without imposing any additional global symmetry, because it is not possible to construct renormalizable or dimension-5 operators that would mediate decays into SM particles \cite{Cirelli:2005uq}. 

We finally remark here that our results also hold  in the case in which  the vev of the new multiplet is  non-zero, provided it is much smaller than the electroweak scale. In this case the global $U(1)$ symmetry is only softly broken and the mixing between the neutral component of the multiplet and the Higgs boson is negligible. For example, this occurs in two-Higgs doublet models with an approximate global $U(1)$ symmetry, where a hierarchy between the vevs of the doublets, $\phi_{1}$ and $\phi_{2}$, can be naturally 
achieved by adding to the scalar potential the term $\mu^{2} \phi_{1}^{\dagger}\phi_{2}$,  with $\mu \ll 1$ GeV, which softly breaks the global symmetry \cite{Ma, Grimus:2009mm}. 
Such explicit breaking of the symmetry can be avoided by the introduction of an additional scalar, say $\phi_{3}$, 
whose vev generates the required term: $\mu^{2}=\mu^{\prime}\langle \phi_{3}\rangle$. As long as $\mu^{\prime} \ll 1$ GeV, the minimum of the scalar potential 
corresponds to $\langle \phi_{2,3}\rangle \propto \mu^{\prime} \ll \langle \phi_{1} \rangle, \langle \phi_{3,2}\rangle$ \cite{Grimus:2009mm} 
(see, $e.g.$, \cite{JosseMichaux:2011ba} for a concrete realization of this scenario).
Similar considerations are valid in the Higgs triplet model, where $S$ is in the adjoint of $SU(2)_{L}$ and has hypercharge $Y=2$.
This scenario is similar to the scalar part of the type II seesaw model for the generation of neutrino masses \cite{typeII}. In this case the global symmetry is equivalent to
the total lepton charge and it is explicitly broken by the term $\mu^{\prime} H^{T} i \tau_{2} S^{\dagger} H$ in the scalar potential. 
After the spontaneous breaking of the electroweak symmetry, the neutral component of the triplet takes a vev, $\langle S^{0} \rangle \simeq \mu^{\prime} v^{2}/2\mu_{S}^{2}\ll v$. 
Indeed, an upper limit on $\langle S^{0} \rangle$ is provided by the measurement of the $\rho$ parameter, 
$\langle S^{0} \rangle /v \lesssim 0.02$ or $\langle S^{0} \rangle \lesssim 5$ GeV \cite{Akeroyd:2012nd}. 
Similarly to the two-Higgs doublet model discussed above, the hierarchy between the vev of the triplet and the electroweak 
scale is possible because the scale $\mu^{\prime}$ is naturally suppressed at all orders, due to the presence of a global (lepton number) symmetry.

The leading order ggZ cross-section for the production of a scalar-antiscalar pair vanishes \cite{delAguila:1990yw}, so that only Higgs-mediation is relevant here. The partonic cross-section with c.o.m. energy $\sqrt{\hat s}$ is
\begin{equation}
	\sigma\left(g\, g \to S^{(Q)}\bar{S}^{(Q)};\,\hat s\right)\;=\;\frac{\Lambda_{Q}^{2}\,G_{H}^{2}\,
	\sqrt{\hat s}\,\sqrt{\hat s-4\,m_{S^{(Q)}}^{2}}}{4096\,\pi\,\left(m_{h}^{2}-\hat s\right)^{2}}\,,\label{sGGF}
\end{equation}
where  the effective coupling $\Lambda_{Q}$ is given in  Eq.~(\ref{LQ}) and the gluon-gluon fusion effective coupling $G_{H}$  is defined in Eq.~(\ref{GHeff}).

\begin{figure}[t!]
\begin{center}
\begin{tabular}{cc}
\includegraphics[width=0.49\textwidth]{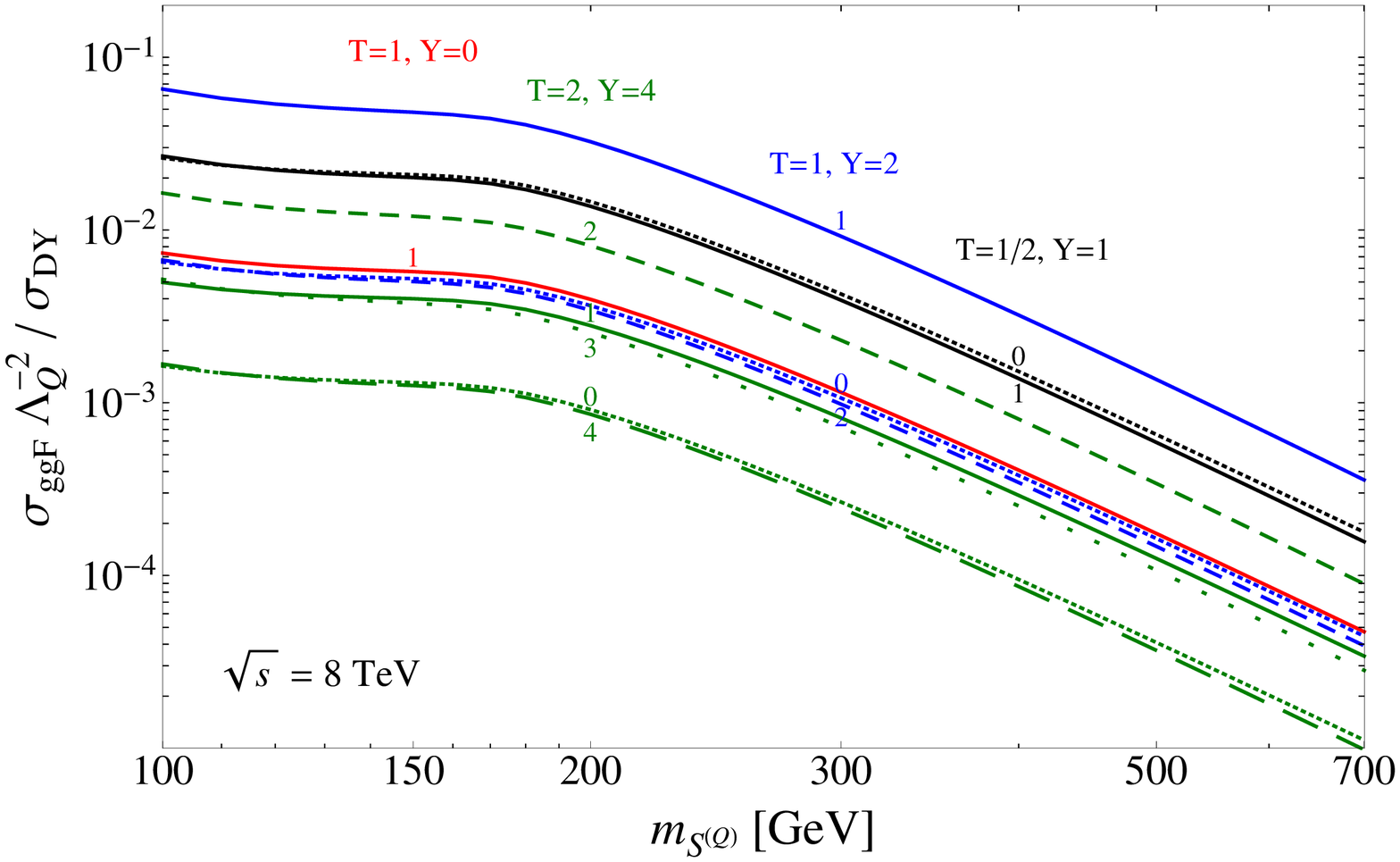} 
\includegraphics[width=0.49\textwidth]{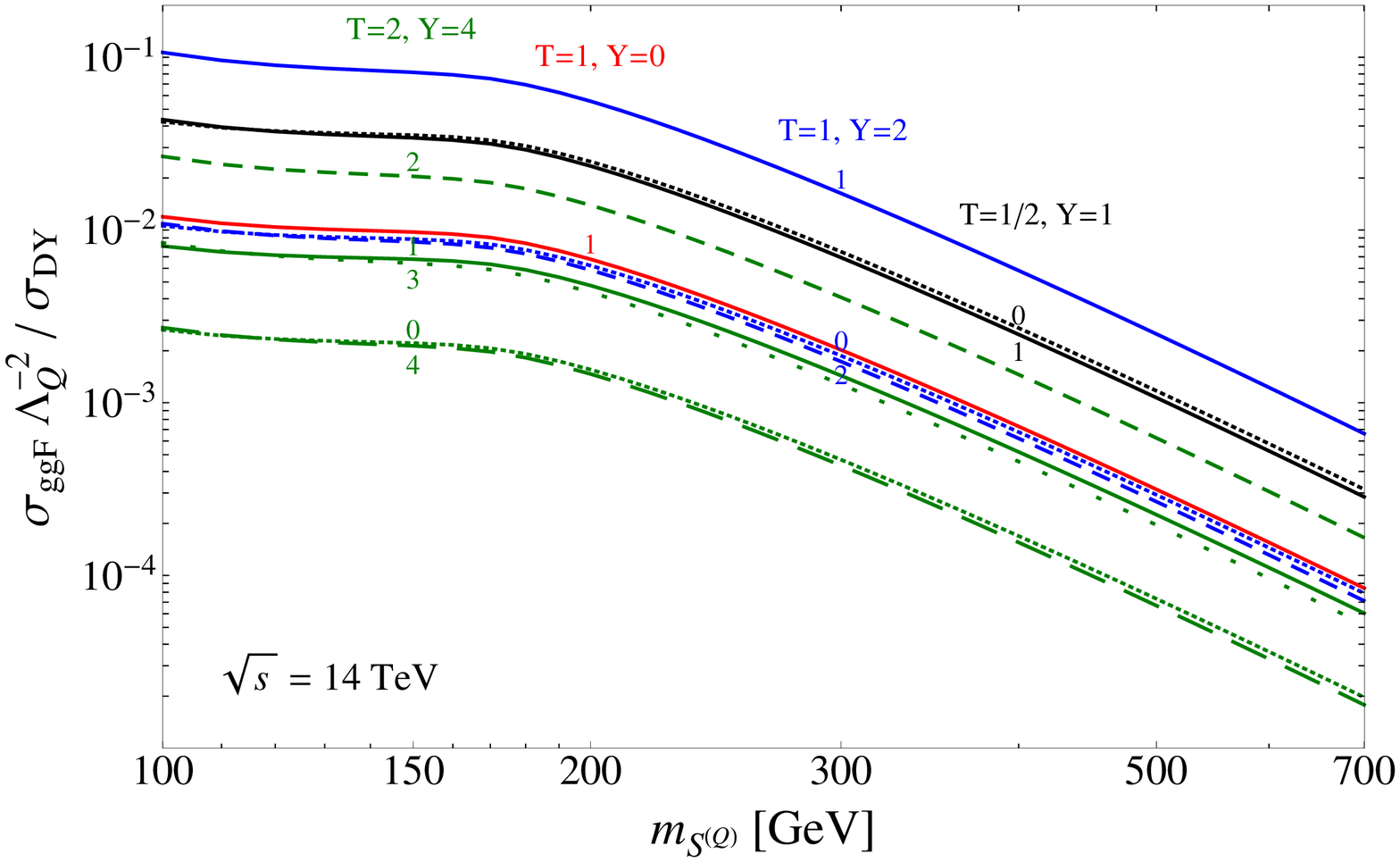}
\end{tabular}
\caption{\small{Relative contribution of the normalized gluon-gluon fusion vertex and the Drell-Yan production cross-sections as function of the scalar mass for 4 different $SU(2)_{L}$ multiplets,
with $\sqrt{s}=8$ TeV (left panel) and 14 TeV (right panel). The number on each line corresponds to the electric charge of the scalar particle.}}
\label{Fig:1}
\end{center}
\end{figure}

On the other hand, the leading order  Drell-Yan production cross-section of a scalar pair for a generic multiplet  of weak isospin $T$ and hypercharge $Y$ reads, for up-antiup quark collisions and neglecting the quark masses,
\begin{eqnarray}
	&&\sigma\left(u\, \bar{u} \to S^{(Q)}\bar{S}^{(Q)};\,\hat s\right) \;= \;\nonumber\\
	&&\frac{\pi \, \alpha_{\rm em}^2 \,\hat s\, \left(1-4 \,m_{S^{(Q)}}^2/\hat s\right)^{3/2}}{288 \,c_w^4 \,s_w^4 \left(m_Z^2-\hat s\right)^2} 
	\left[4\, Q^2 \,c_w^4   \left(\frac{16\, m_Z^4 s_w^4}{9\, \hat s^2}-1\right)+\,4 \left(Q  \, c_w^2-\frac{2\, Y\, s_w^2}{3}\right)^2\right.\nonumber\\
   &&\left.+\left(2 \,Q \,c_w^2 \left(1-\frac{4\, m_Z^2
   s_w^2}{3\, \hat s}\right)-Y \left(1-\frac{8\, s_w^2}{3}\right)\right)^2+\frac{8}{3} \,Y^2 \,s_w^2
   \left(c_w^2-s_w^2\right)\right]\,,\label{sDYu}
\end{eqnarray}
while for the down-antidown quark initial state,
\begin{eqnarray}   
   &&\sigma\left(d\, \bar{d} \to S^{(Q)}\bar{S}^{(Q)};\,\hat s\right) \;= \;\nonumber\\
	&&\frac{\pi  \,\alpha_{\rm em}^2\, \hat s\, \left(1-4 \,m_{S^{(Q)}}^2/\hat s\right)^{3/2}}{288 \,c_w^4 \,s_w^4 \left(m_Z^2-\hat s\right)^2} 
	\left[4\, Q^2 \,c_w^4   \left(\frac{4\, m_Z^4 s_w^4}{9\, \hat s^2}-1\right)+\,4 \left(Q  \, c_w^2-\frac{Y\, s_w^2}{3}\right)^2\right.\nonumber\\
   &&\left.+\left(2 \,Q \,c_w^2 \left(1-\frac{2\, m_Z^2
   s_w^2}{3\, \hat s}\right)-Y \left(1-\frac{4\, s_w^2}{3}\right)\right)^2+\frac{4}{3} \,Y^2 \,s_w^2
  \,c_w^2\right]\,.\label{sDYd}
\end{eqnarray}
 For charged scalar production, there is a contribution to the production cross-section both from photon and from $Z$ boson exchange, hence the DY partonic cross-sections can be  relatively enhanced or suppressed depending on the electric charge and weak isospin of each multiplet component.  Notice that for $Y=0$ and odd $n$ the neutral component of the multiplet cannot be pair produced through electroweak interactions.
The total production cross-section at the LHC is obtained  by convoluting the single partonic cross-sections with the appropriate parton distribution functions, as described in section~\ref{Sec:Basics}.\footnote{The numerical results have been checked against a computation using CalcHEP 3.4  \cite{Belyaev:2012qa}.}

\begin{figure}[t!]
\begin{center}
\begin{tabular}{cc}
\includegraphics[width=0.48\textwidth]{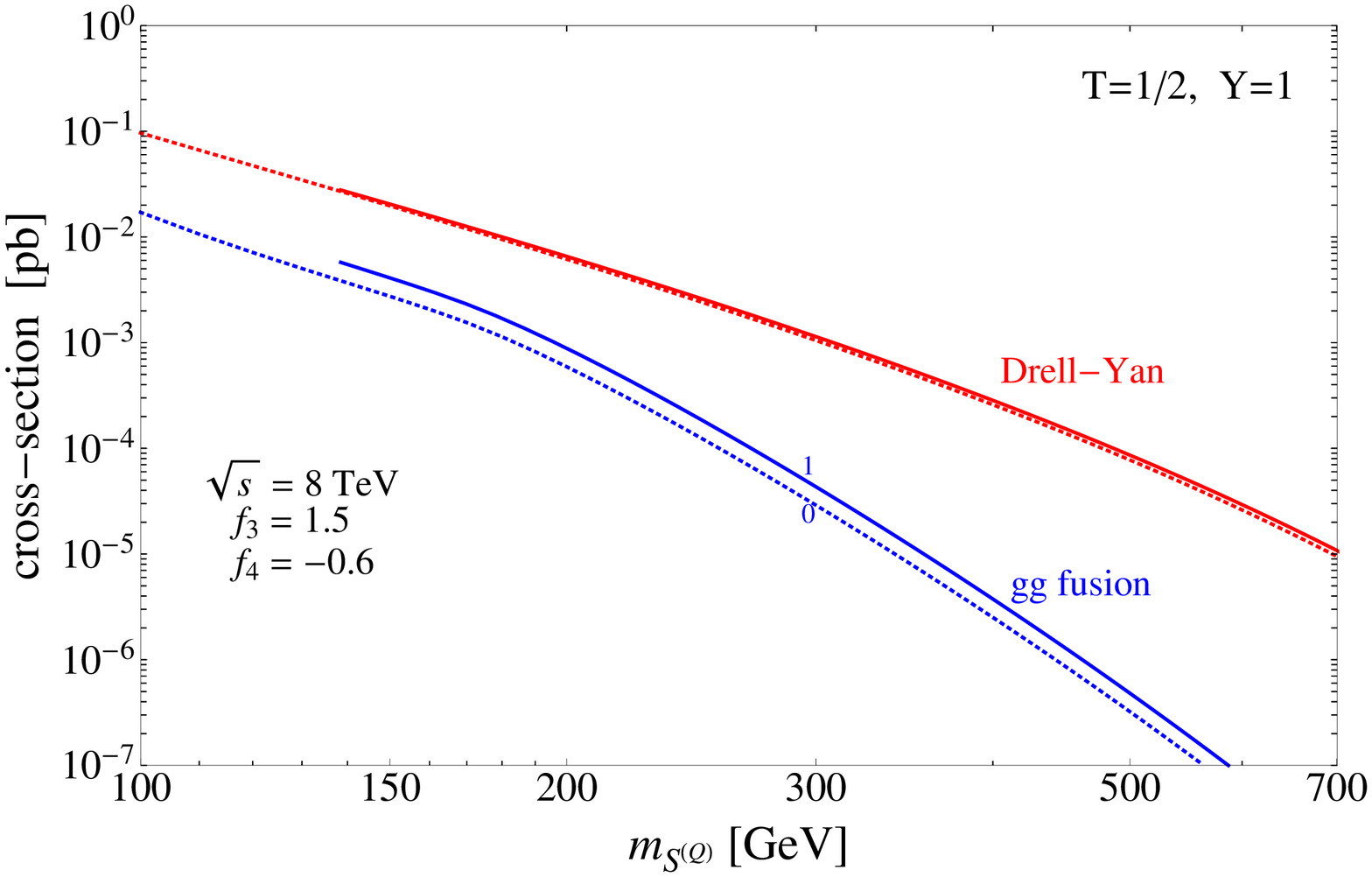} 
\includegraphics[width=0.48\textwidth]{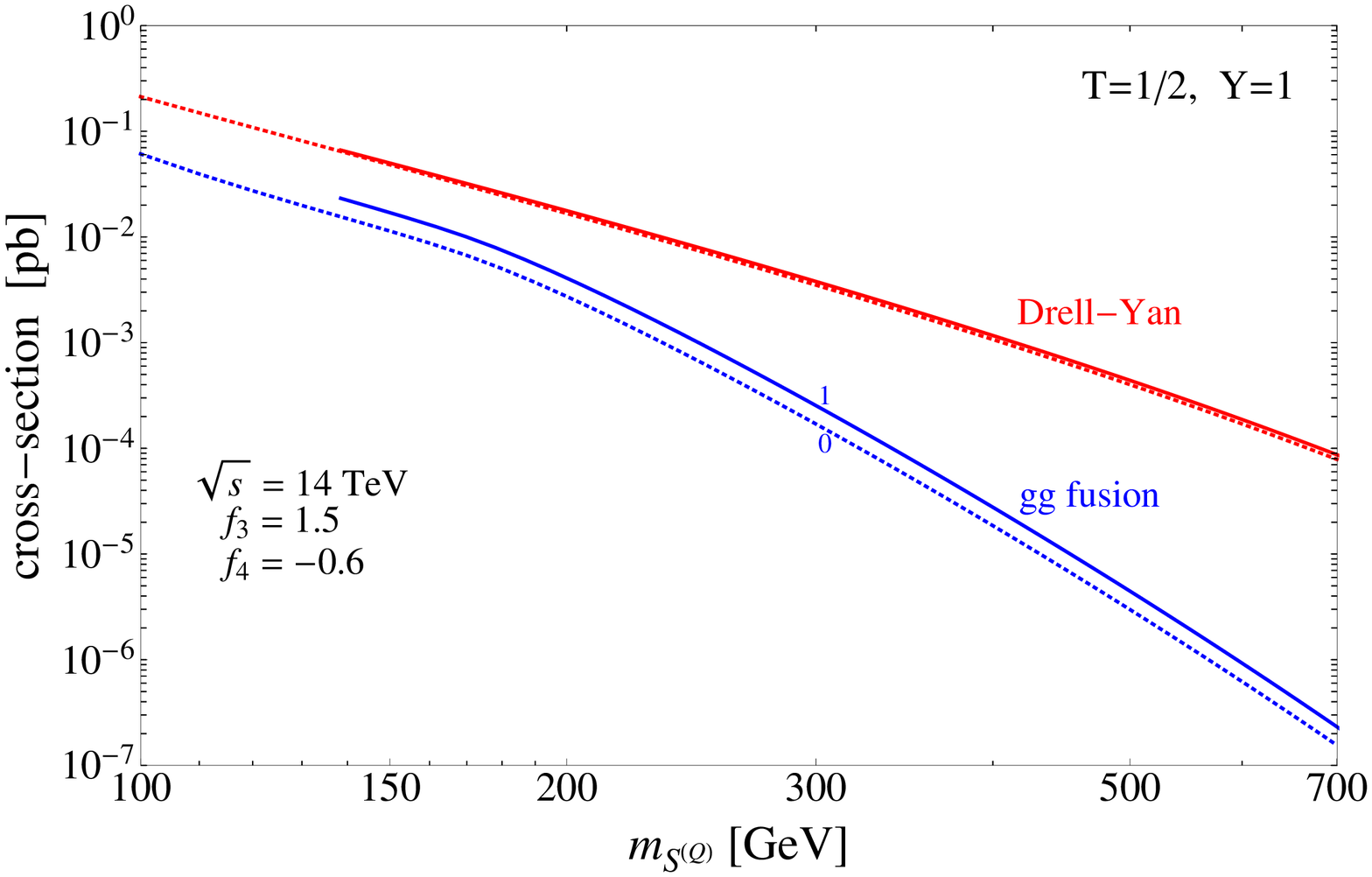}\\ 
\includegraphics[width=0.48\textwidth]{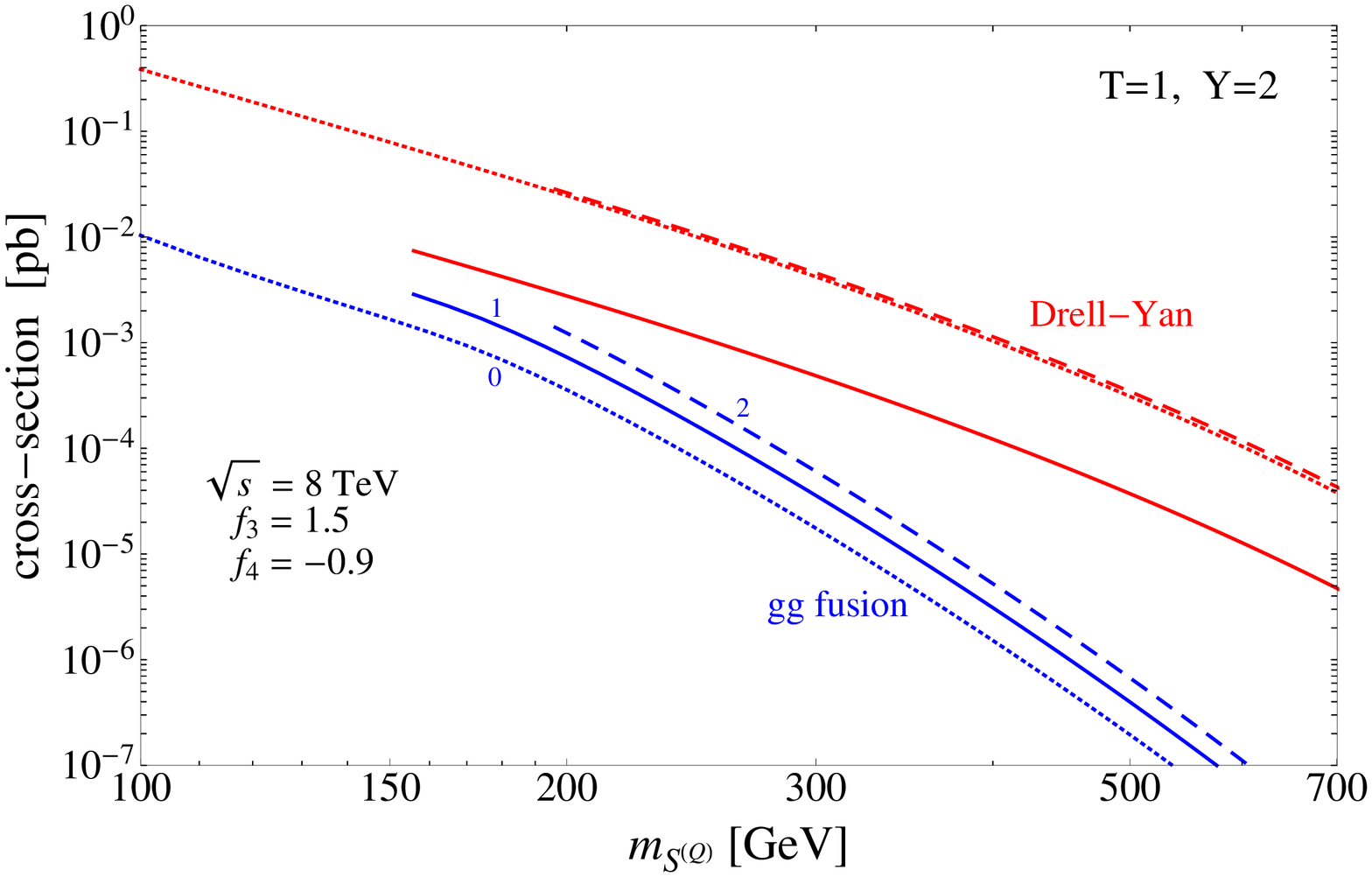} 
\includegraphics[width=0.48\textwidth]{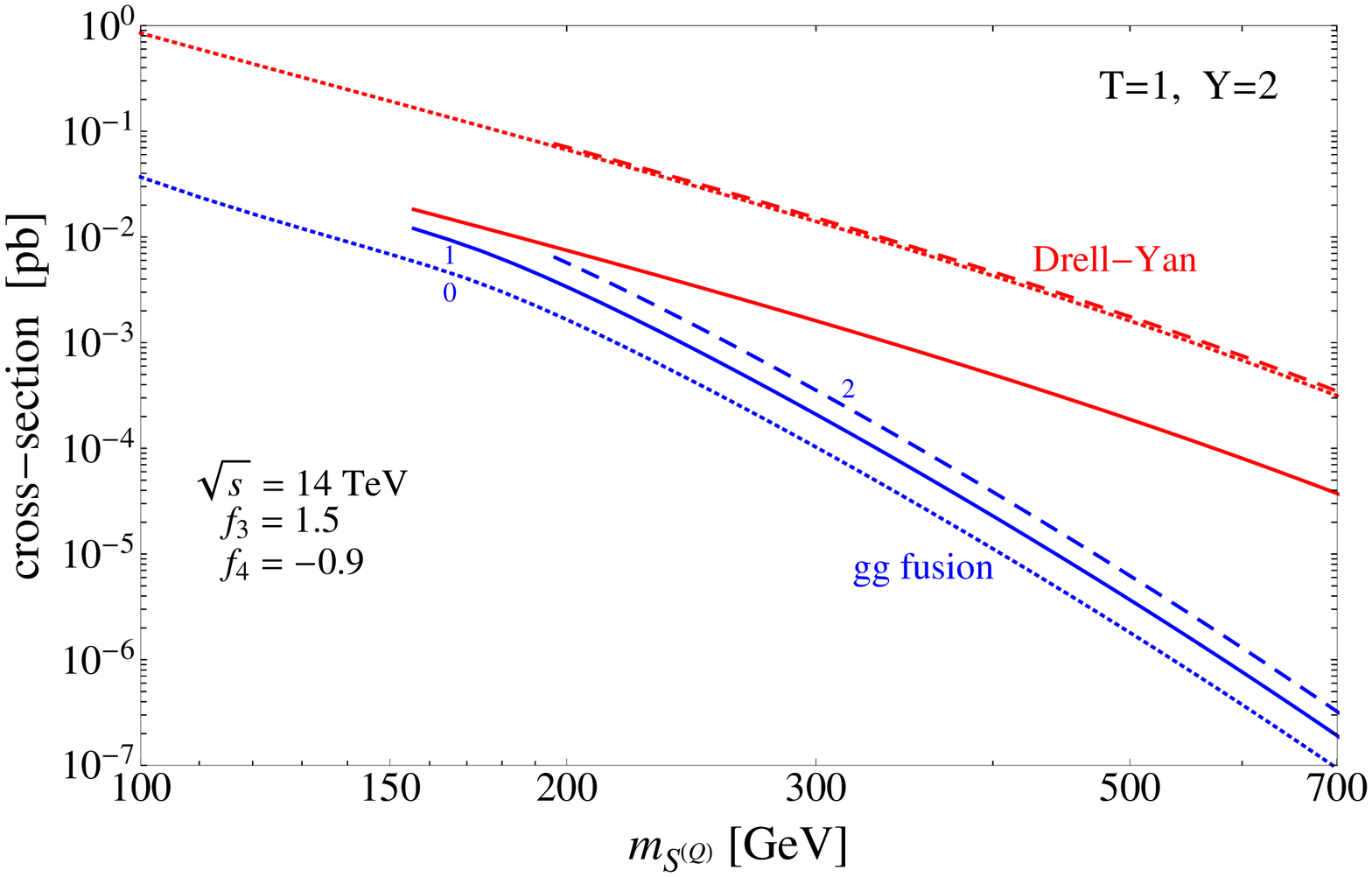} 
\end{tabular}
\caption{\small{Gluon-gluon fusion and Drell-Yan production cross-sections for two different representations of $SU(2)_L$. The dashed, solid and dotted lines correspond to the production of scalars with electric charge equal to $+2$, $+1$ and $0$, respectively. For the doublet, the effective couplings are $\Lambda_{+1}=-3.3$, $\Lambda_{0}=-2.7$ and, for the triplet, $\Lambda_{+2}=-3.9$, $\Lambda_{+1}=-3$ and $\Lambda_{0}=-2.1$. }}
\label{Fig:2}
\end{center}
\end{figure}

We show in Fig.~\ref{Fig:1} the ratio between the gluon-gluon fusion cross-section $\sigma_{\rm ggF}$ normalized to the squared effective scalar coupling $\Lambda_{Q}$ and the Drell-Yan  production cross-section $\sigma_{\rm DY}$ for proton center-of-mass energies of 8 TeV (left panel) and 14 TeV (right panel).
 This quantity is, for a fixed $m_{S^{(Q)}}$, independent of the quartic couplings in the scalar potential and is determined exclusively by the electroweak quantum numbers of the scalar multiplet, as can be checked from Eqs. (\ref{sGGF}-\ref{sDYd}).
  The overall cross-section for each production mechanism is calculated using Eqs.~(\ref{DYCS}-\ref{lum2})~and~(\ref{sGGF}-\ref{sDYd}) for several $SU(2)_{L}$ representations with fixed hypercharge. The numbers on each line represent the electric charge of the scalar mass eigenstate. In the numerical computation we use the CTEQ6 PDFs \cite{Pumplin:2002vw} and we fix both the factorization scale $\mu_{F}$ and the renormalization scale of the strong coupling constant  at the invariant mass of the final state. 
As is noticeable from the plot, the dependence of the Higgs-mediated production cross-section with the scalar mass  changes when $m_{S^{(Q)}}\simeq m_t$, which is due to the different functional form of the Higgs coupling with gluons when the partonic center-of-mass energy is larger or smaller than the top mass, see Eq. (\ref{loopfunction}). Moreover, we find that for the electrically neutral component of the multiplet, the ratio $\sigma_{\rm ggF}\Lambda^{-2}_Q/\sigma_{\rm DY}$ decreases as the isospin increases. Nevertheless, due to the complicated dependence of the Drell-Yan cross-section with the electric charge and the hypercharge, it is not possible to ascertain a pattern within the members of the same multiplet.

More specifically, we show in Fig.~\ref{Fig:2} the scalar production cross-sections associated with the ggF (blue lines) and DY (red lines) sub-processes for fixed values of the quartic couplings $f_{3}$ and $f_{4}$.
We consider two specific scenarios: $i)$ doublet with $Y=1$ (upper panels) and $ii)$ triplet with $Y=2$ (lower panels). We have checked 
that for this choice of quantum numbers the couplings are in agreement with  the constraints on the parameter space discussed earlier in this section. 
The dashed, solid and dotted lines correspond to the production of scalars with electric charge equal to $+2$, $+1$ and $0$, respectively. 
Notice that Eq.~(\ref{eq:mQ}) sets a lower bound on the mass of each component of the multiplet, which has been taken into account for the mass range depicted in Fig.~\ref{Fig:2}. 
In the  shown mass range, the
$h \rightarrow \gamma \gamma$ signal strength is consistent with the CMS measurement~\cite{CMS:2013wda} at the  $2 \sigma$ level, while the $T$ parameter does not deviate from the best fit value~\cite{Baak:2014ora} by more than $1 \sigma$.
We find, in particular, that the production cross-section of the neutral and charged component of the doublet is approximately a factor 5 (2) smaller than the corresponding Drell-Yan cross-section for $m_{S^{(Q)}}\simeq 140$ GeV and  c.o.m. energy of 8 (14) TeV.   For the scalar triplet, the gluon-gluon fusion and Drell-Yan production cross-sections of the singly charged component differ by a factor smaller than 4 (2) for $m_{S^{(+1)}}\lesssim 180 $ GeV at 8 (14) TeV, while electroweak interactions provide clearly the dominant contribution to the production of the neutral and doubly charged scalars, within all the mass range.

\section{Fermion production at the LHC}
\label{Sec:fermion}

We consider in this section the impact of the gluon-gluon fusion mechanism on the production of new fermionic representations at the LHC. 
In what follows we assume  a simplified scenario with just one extra uncoloured fermion multiplet $\psi$, which has Yukawa-type interactions
involving the Higgs doublet and the SM leptons $L$, $E$:~\footnote{A simple interesting alternative  are models with vector-like fermions, see e.g.  \cite{Delgado:2011iz} and \cite{Ma:2014zda}. For an analysis of the impact of gluon-gluon fusion in these models, see  \cite{Liu:2010ze}.}
\begin{equation}
{\mathcal{L}}_\psi \; = \; c\bar{\psi}\left(i\slashed{D}-m_\psi\right)\psi + {\mathcal{L}}_{\mathrm{Y}}\left(H,\psi,L,E\right)\,,\label{Lpsi}
\end{equation}
where $c$ is a constant which equals 1 (1/2) if $\psi$ is a Dirac (Majorana) fermion. Depending on the quantum number assignments, such an interaction may or may not exist at the renormalizable level. The spontaneous breaking of the electroweak symmetry results in an interaction between the Higgs boson and the new fermionic mass eigenstate(s),  
as well as a mixing between SM lepton flavours and the new fermion(s), 
which can be sizeable depending on the details of the model. The requirement of renormalizability of Lagrangian 
(\ref{Lpsi}) restricts our analysis to the case in which the new fermion $\psi$ has zero hyperchange and is either a singlet or a triplet of $SU(2)_L$, as in the well known type I \cite{typeI} and type III \cite{typeIII} seesaw extensions of the SM. Here, we have additional heavy right-handed Majorana neutrinos which mix with the SM left-handed neutrinos. The mixing  between the heavy and light neutrinos can be encoded in a matrix $\Theta_{\alpha k}$, which enters the expression of the light neutrino mass matrix, $(m_{\nu})_{\alpha \beta} = \Theta_{\alpha k}^{*} \,M_{k}\, \Theta^{\dagger}_{k\beta}$.  The matrix elements of $\Theta_{\alpha k}$ are constrained by lepton flavour violating processes, notably by $\mu\to e\,\gamma$. Typically, $\left|\Theta_{\alpha k}\right|\lesssim 10^{-2}$ in the type I seesaw for a heavy neutrino mass $M\sim 100$ GeV, whereas the type III requires smaller values  for $\left|\Theta_{\alpha k}\right|$ \cite{TeVseesaw}.

We discuss in this  framework the production of heavy Majorana fermions, $\mathcal{N}_{k}$, and charged Dirac fermions, $\mathcal{E}_{k}$, with a mass $M_{k}$ varying in the  electroweak range (see, $e.g.$, \cite{TeVseesaw} for a phenomenological analysis of such seesaw scenarios). The production proceeds either through the Drell-Yan processes or through the gluon-gluon fusion mechanism involving a Higgs and a $Z$, as discussed in section~\ref{Sec:Basics}. For a collider signal analysis of the far stronger channels $q\bar{q}'\rightarrow W^*\rightarrow\mathcal{N}\ell^\pm$ for type I and $q\bar{q}\rightarrow Z^*\rightarrow\mathcal{E}^+\mathcal{E}^-$, $q\bar{q}'\rightarrow W^*\rightarrow \mathcal{E}^\pm\mathcal{N}$ for type III, see, $e.g.$, \cite{delAguila:2008cj}. These channels are not considered here because either the final state cannot be obtained via a Higgs-mediated process, or it is suppressed by a higher power of the mixing parameter.

In the particular case of heavy Majorana neutrinos, $\mathcal{N}_{k}$ ($k=1,2,\ldots$), involved in the generation of active neutrino masses through the type I/III seesaw mechanism, the relevant couplings to the $Z$ and the Higgs boson
 after electroweak symmetry breaking can be parametrized by the  following effective Lagrangian terms
\begin{eqnarray}
\mathcal{L}_{NC}^\mathcal{N} &=& -\frac{g}{4 \,c_{w}}\,
\overline{\nu_{\alpha L}}\,\gamma_{\mu}\,\Theta_{\alpha k}\,(1 - \gamma_5)\,\mathcal{N}_{k}\,Z^{\mu}\;
+\; {\rm h.c.}\,,\label{NNC}\\
\mathcal{L}_{H}^\mathcal{N} &=& -\frac{g \,M_{k}}{4\, m_{W}}\,
\overline{\nu_{\alpha L}}\,\Theta_{\alpha k}\,(1 + \gamma_5)\,\mathcal{N}_{k}\,h\;
+\; {\rm h.c.}\,,
\label{NH}
\end{eqnarray}
where $M_{k}$ is the mass of the $k$th heavy Majorana neutrino and $m_{W}$ denotes the $W$ boson mass. 
 
 In the type III seesaw scenario, the SM Lagrangian is extended with new fermionic representations in the adjoint of the weak gauge group. As a consequence, there exists for each  Majorana neutrino $\mathcal{N}_{k}$  one Dirac fermion $\mathcal{E}_{k}$ with electric charge $-1$ and degenerate in mass with $\mathcal{N}_{k}$ at leading order. These new heavy charged fermions  are coupled to the electroweak gauge bosons and the SM leptons through the mixing matrix $\Theta_{\alpha k}$:
 \begin{eqnarray}
\mathcal{L}_{NC}^{\mathcal{E}} &=&
-\,\frac{g}{2\,\sqrt{2}\, c_{w}}\,\overline{\ell_{\alpha}}\,\gamma_{\mu}\,\Theta_{\alpha k}(1-\gamma_{5})\,\mathcal{E}_{k}\,
Z^{\mu}\;+\;{\rm h.c.}\label{ENC}
\end{eqnarray}
Similarly, the coupling of $\mathcal{E}_{j}$ and the Higgs boson is parametrized by the interaction Lagrangian
\begin{eqnarray}
\mathcal{L}_{H}^{\mathcal{E}} &=& -\frac{g \,M_{k}}{2\,\sqrt{2}\, m_{W}}\,
\overline{\ell_{\alpha}}\,\Theta_{\alpha k}\,(1 + \gamma_5)\,\mathcal{E}_{k}\,h\;
+\; {\rm h.c.}\label{EH}
\end{eqnarray}

We consider now the possibility of producing heavy fermionic states in the framework of the low-scale seesaw scenario. The relevant interactions for the production through the ggH portal are given in  Eqs.~(\ref{NH}) and (\ref{EH}). Away from the Higgs resonance and neglecting the mass of the SM leptons, we obtain for the production cross-section via an intermediate Higgs boson:
\begin{equation}
	 \sigma_H\left(g\, g \to \mathcal{N}_{k}\,\nu_{\alpha}\,;\,\hat s\right)\;=\;
	      \sigma_H\left(g\, g \to \mathcal{E}_{k}\,\overline{\ell_{\alpha}}\,;\,\hat{s}\right)
	\;=\;\frac{\pi\,\alpha_{\rm em}^2\,\left|\Theta_{\alpha k}\right|^{2}\,G_{H}^{2}\,M_{k}^{2}\,\left(M_{k}^{2}-\hat{s}\right)^{2}}{512\,s_{w}^{4}\,c_w^4\,m_{Z}^{2}\,\left(m_{h}^{2}-\hat{s}\right)^{2}}\,,
	\label{ggFfer}
\end{equation}
where the effective gluon-gluon-Higgs coupling $G_{H}$ is given in Eq.~(\ref{GHeff}). Notice that, for fermion masses $M_{k}$ below the Higgs boson mass, the corresponding $\mathcal{N}_{k}$ or $\mathcal{E}_{k}$ production cross-section is resonantly enhanced.

On the other hand, the relevant interactions for the neutral current interactions are given in 
Eqs.~(\ref{NNC}) and (\ref{ENC}). The partonic production cross-section of one heavy fermion and one massless SM lepton via the ggZ interaction reads
\begin{eqnarray}
\sigma_Z\left(g\, g \to \mathcal{N}_{k}\,\nu_{\alpha}\,;\,\hat s\right) & = &
\sigma_Z\left(g\, g \to \mathcal{E}_{k}\,\ell_{\alpha}\,;\,\hat s\right)\nonumber\\
& = &
\frac{\alpha_\mathrm{em}\,\left|\Theta_{\alpha k}\right|^2\,G_Z^2\,M_k^2\,\hat s^2\,\left(1-\frac{M_k^2}{\hat{s}}\right)^2}{2048\,s_w^2\,c_w^2\,m_Z^4}\,,
\end{eqnarray}
where $G_Z$ is given in Eq. \eqref{eq:GZ}. As mentioned in section \ref{Sec:Basics}, we only consider the top quark contribution. Note also the absence of a resonance in $\hat{s}=m_Z^2$ due to the Landau-Yang theorem \cite{LandauYang:19481950}: an on-shell $Z$ boson cannot be produced in the collision of two massless spin-1 vector bosons.

Lastly, the expressions for the production cross-section in a quark-antiquark collision read:
\begin{eqnarray}
	\sigma\left(u\,\overline{u}\to \mathcal{N}_{k}\,\nu_{\alpha}\,;\,\hat{s}\right) &=&
	\sigma\left(u\,\overline{u}\to \mathcal{E}_{k}\,\overline{\ell_{\alpha}}\,;\,\hat{s}\right)\\
	&=&  \frac{\pi\,\alpha_{\rm em}^{2}\left|\Theta_{\alpha k}\right|^{2}\left(\hat{s}-M_{k}^{2}\right)^{2}\left(M_{k}^{2}+2\,\hat{s}\right)}{1296\,c_{w}^{4}\,s_{w}^{4}
          \left(\hat{s}-m_{Z}^{2}\right)^{2}\,\hat{s}^{2}}\left(9\,c_{w}^{4}-6\,c_{w}^{2}\,s_{w}^{2}+17\,s_{w}^{4}\right),\nonumber\\\nonumber\\
        \sigma\left(d\,\overline{d}\to \mathcal{N}_{k}\,\nu_{\alpha}\,;\,\hat{s}\right) &=&
          	\sigma\left(d\,\overline{d}\to \mathcal{E}_{k}\,\overline{\ell_{\alpha}}\,;\,\hat{s}\right)\label{DYfer}\\
	 &=& \frac{\pi\,\alpha_{\rm em}^{2}\left|\Theta_{\alpha k}\right|^{2}\left(\hat{s}-M_{k}^{2}\right)^{2}\left(M_{k}^{2}+2\,\hat{s}\right)}{1296\,c_{w}^{4}\,s_{w}^{4}
          \left(\hat{s}-m_{Z}^{2}\right)^{2}\,\hat{s}^{2}}\left(9\,c_{w}^{4}+6\,c_{w}^{2}\,s_{w}^{2}+5\,s_{w}^{4}\right).\nonumber
\end{eqnarray}
The total fermion production cross-section at the LHC, for a c.o.m. energy $\sqrt{s}$, is given by  the convolution of the corresponding cross-section, Eqs.~(\ref{ggFfer}-\ref{DYfer}), with the corresponding parton luminosity functions, as described in Eqs.~(\ref{DYCS}-\ref{lum2}) and taking $\tau_{s}\equiv M_{k}^{2}/s$ in the limit of massless charged leptons.

\begin{figure}[t!]
\begin{center}
\begin{tabular}{cc}
\includegraphics[width=0.49\textwidth]{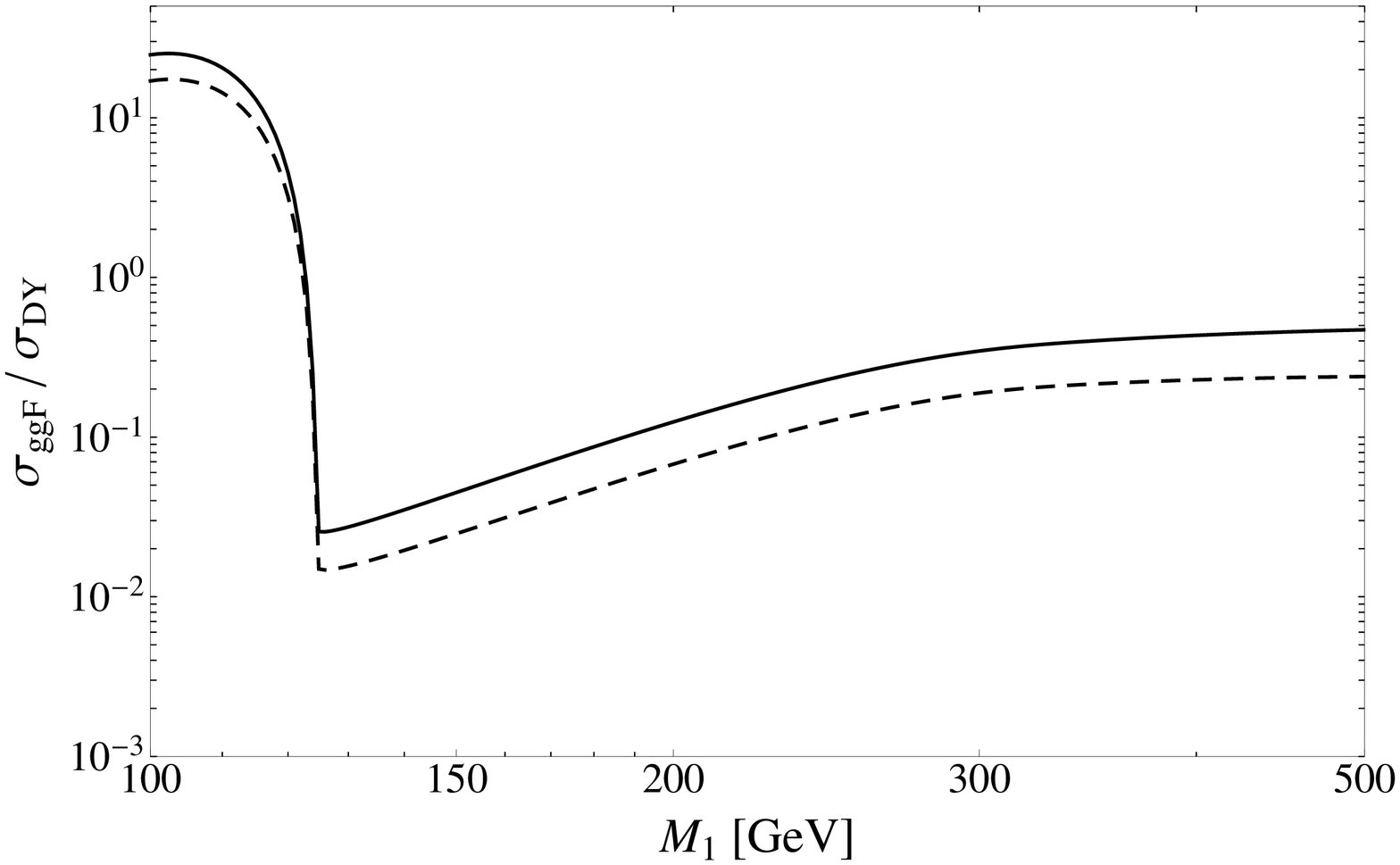} 
\includegraphics[width=0.49\textwidth]{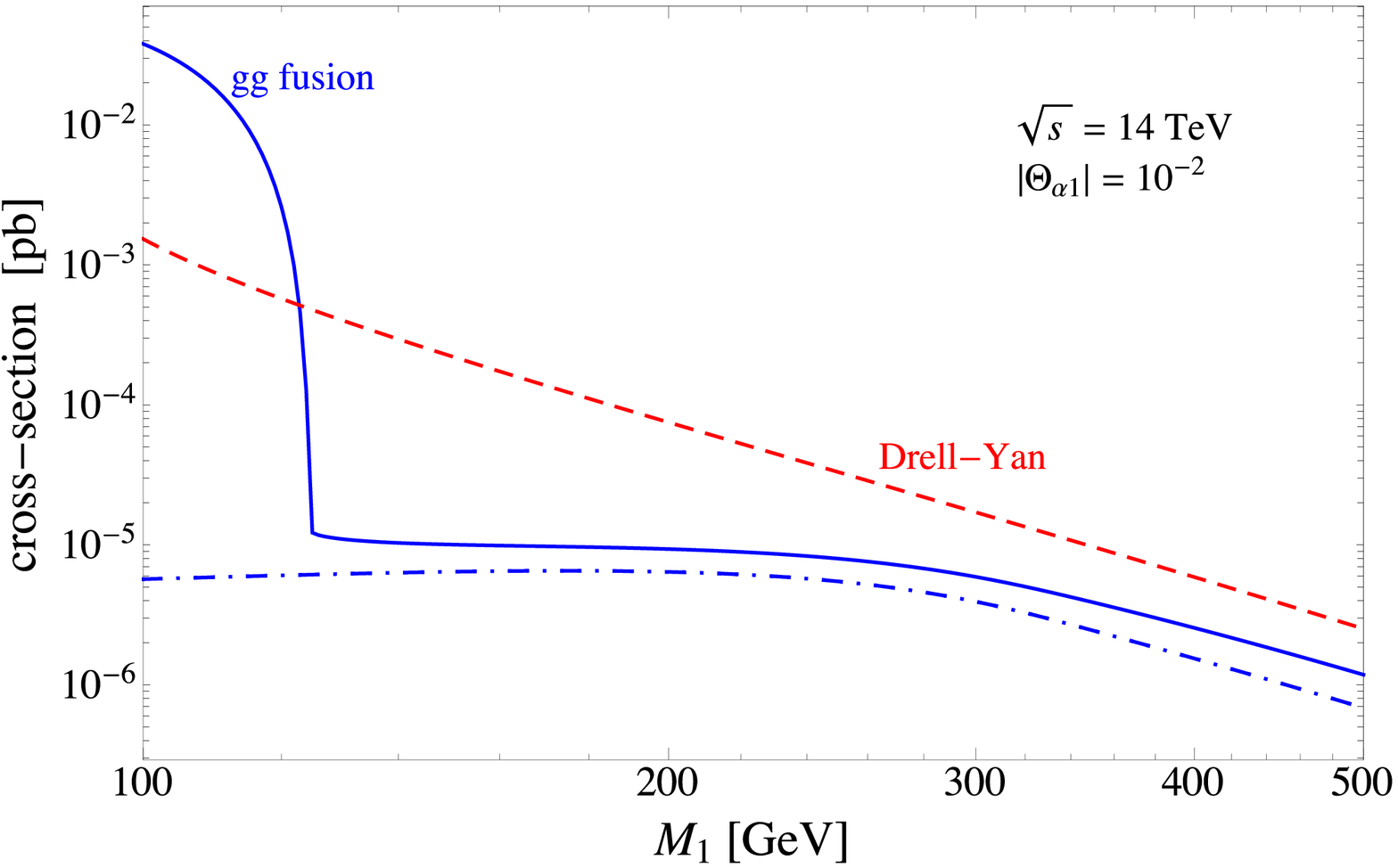}
\end{tabular}
\caption{\small{{\it Left panel:} ratio between gluon-gluon fusion and Drell-Yan production cross-sections of heavy fermions in type I/III seesaw scenarios at $\sqrt{s} = $8 (14) TeV, dashed (continuous) line. {\it Right panel:} Fermion production cross-sections for $\sqrt{s} = $14 TeV and $|\Theta_{\alpha1}|=10^{-2}$. The solid blue line represents the complete contribution due to gluon-gluon fusion, while the dot-dashed blue line shows the contribution from ggZ alone.}}
\label{Fig:3}
\end{center}
\end{figure}

We report in Fig.~\ref{Fig:3}, left panel, the ratio between the gluon-gluon fusion and the Drell-Yan production cross-sections of a Majorana neutrino $\mathcal{N}_{1}$ or charged fermion $\mathcal{E}_{1}$ as a function of the fermion mass $M_{1}$. The two curves correspond to the LHC c.o.m. energy of 8 TeV (dashed line) and 14 TeV (continuous line), respectively.  Notice that, at lowest order in the mixing parameter,  the ratio of $\sigma_{\rm ggF}$ and $\sigma_{\rm DY}$ for these final states does not depend on the mixing between heavy fermions and SM leptons. We also report  on the right panel the production cross-sections for a fixed value of the mixing parameter, $|\Theta_{\alpha1}|=10^{-2}$, and $\sqrt{s}=14$ TeV. As apparent from the plot, the (Higgs-mediated) gluon-gluon fusion vertex provides the largest contribution to the  fermion production cross-section for masses  $M_{k}\lesssim m_h$, due to the resonant enhancement for fermion masses below the Higgs pole. This opens the possibility of testing the type I scenario at the energy frontier via Higgs boson exotic decays. For example, the production and detection of the heavy Majorana fermions might be feasible via the chain $h\to \nu_{\alpha L}\, N_{j}$, with $N_j\rightarrow \ell_{\beta}\,W, \nu Z$ and the gauge bosons subsequently decaying producing  jets or leptons (see \cite{Cely:2012bz,BhupalDev:2012zg} for detailed discussions about these signals at the LHC).
On the other hand, for larger heavy fermion masses the gluon-gluon fusion process only gives a subdominant contribution to the total production cross-section, although this contribution can be as large as 40\%-50\% for $\sqrt{s}=14$ TeV and should not be neglected. This is the case, in particular, of the  type III seesaw, where the extra fermion masses are restricted to be $m_{\mathcal{N}} \gtrsim 300$ GeV, as follows from the ATLAS search  for the final state $\mathcal{E N}$ presented in \cite{ATLAS:2013hma}.

\mathversion{bold}
\section{Perspectives for a $\sqrt{s}= 100$ TeV proton-proton collider}
\label{Sec:100TeV}
\mathversion{normal}

There is currently a growing discussion within the Particle Physics community regarding the first steps towards the post-LHC era. In particular, the planning of future high energy facilities is a long term endeavour and requires careful consideration of the expected physics opportunities. In this context, it is timely to assess the impact of the gluon-gluon fusion mechanism on the production rate of new particles beyond the SM at future machines, such as a high energy upgrade of the LHC with 33 TeV center-of-mass energy \cite{Todesco:2011np} or a hypothetical 100 TeV  proton-proton collider \cite{100TeVBSM}. In the following, we will concentrate our discussion mostly on the latter possibility.

\begin{figure}[t!]
\begin{center}
\begin{tabular}{cc}
\includegraphics[width=0.48\textwidth]{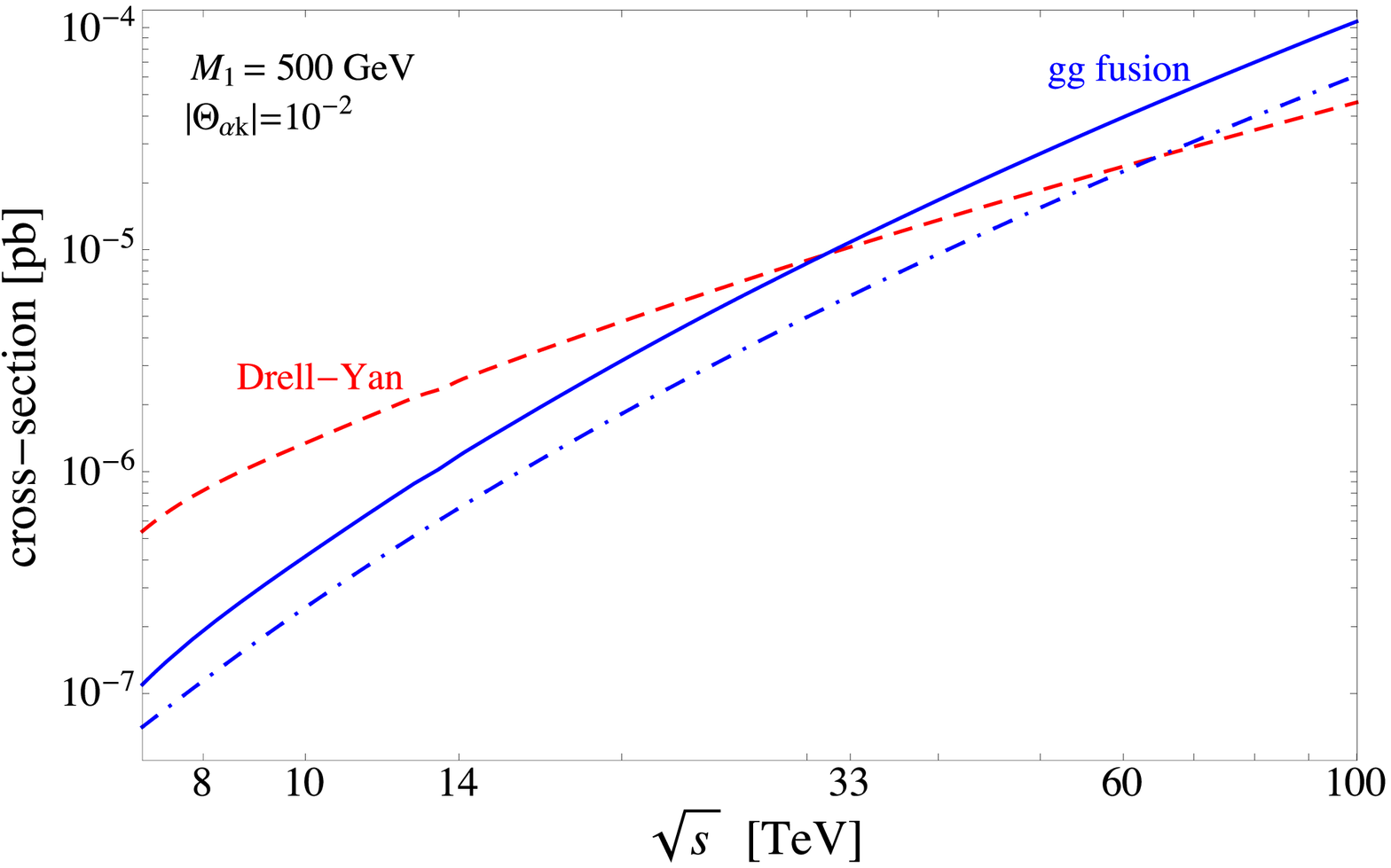} 
\includegraphics[width=0.48\textwidth]{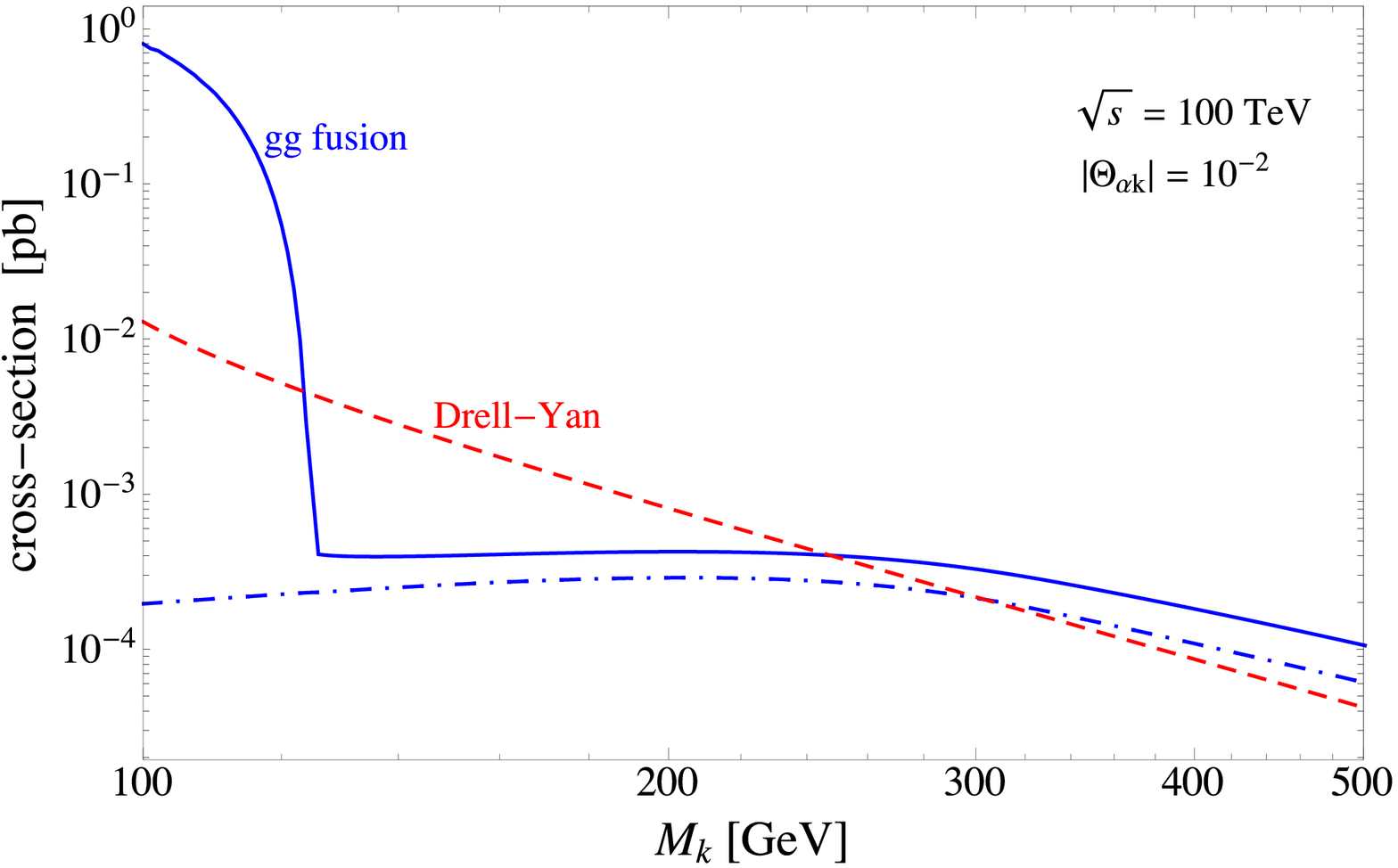}\\ 
\end{tabular}
\caption{\small{{\it Left panel}: Cross-section for the production of an $\mathcal{E}_k \, \ell_\alpha$ or $\mathcal{N}_k \, \nu_\alpha$ pair with $|\Theta_{\alpha k}| = 10^{-2}$ both from Drell-Yan and gluon-gluon fusion for varying $\sqrt{s}$. {\it Right panel}: Fermion production cross-sections for $\sqrt{s} = $100 TeV and $|\Theta_{\alpha1}|=10^{-2}$ as a function of $M_k$.
}}
\label{Fig:sqrts}
\end{center}
\end{figure}

We consider for concreteness
the production of a charged component of a  fermionic triplet $\mathcal{E}_k$ with mass $M_k = 500 \GeV$ and a mixing parameter $|\Theta_{\alpha k} | = 10^{-2}$, in association with a SM fermion $\ell_\alpha$. As can be seen in Fig.~\ref{Fig:sqrts},  left panel, a higher center-of-mass energy has two main effects on the cross-section. First, a larger value of $\sqrt{s}$ at a hadron collider  implies a larger cross-section for both Drell-Yan and the gluon-gluon fusion mechanism, since parton luminosities increase and a larger  multi-body phase space becomes available. Second, the relative importance of different partons changes, thus leading to both  more flavour democratic contributions to Drell-Yan as well as an increasing relevance of gluon initial states.
 For this exemplary case, the relative importance of the gluon-gluon fusion increases by one order of magnitude  and hence the gluon-gluon fusion cross-section $\sigma_{\rm ggF} $ becomes as important as the Drell-Yan contribution $\sigma_{\rm DY}$ at $\sqrt{s} = 30 \TeV$ and is completely dominant at $\sqrt{s} = 100 \TeV$, as shown in Fig.~\ref{Fig:sqrts} (left panel). 
More specifically, and as apparent from Fig.~\ref{Fig:sqrts} (right panel), for a 100 TeV proton-proton collider the gluon-gluon fusion contribution is the dominant production channel for the range of masses currently allowed by the ATLAS search of the charged component of the fermionic triplet \cite{ATLAS:2013hma}, $M_k\gtrsim 300$ GeV.
Interestingly,   with a proposed luminosity of up to $L = 10 \; \mbox{ab}^{-1}$,  hundreds of events can be expected,  provided $M_k\lesssim 500$ GeV.
\begin{figure}[t!]
\begin{center}
\begin{tabular}{cc}
\includegraphics[width=0.48\textwidth]{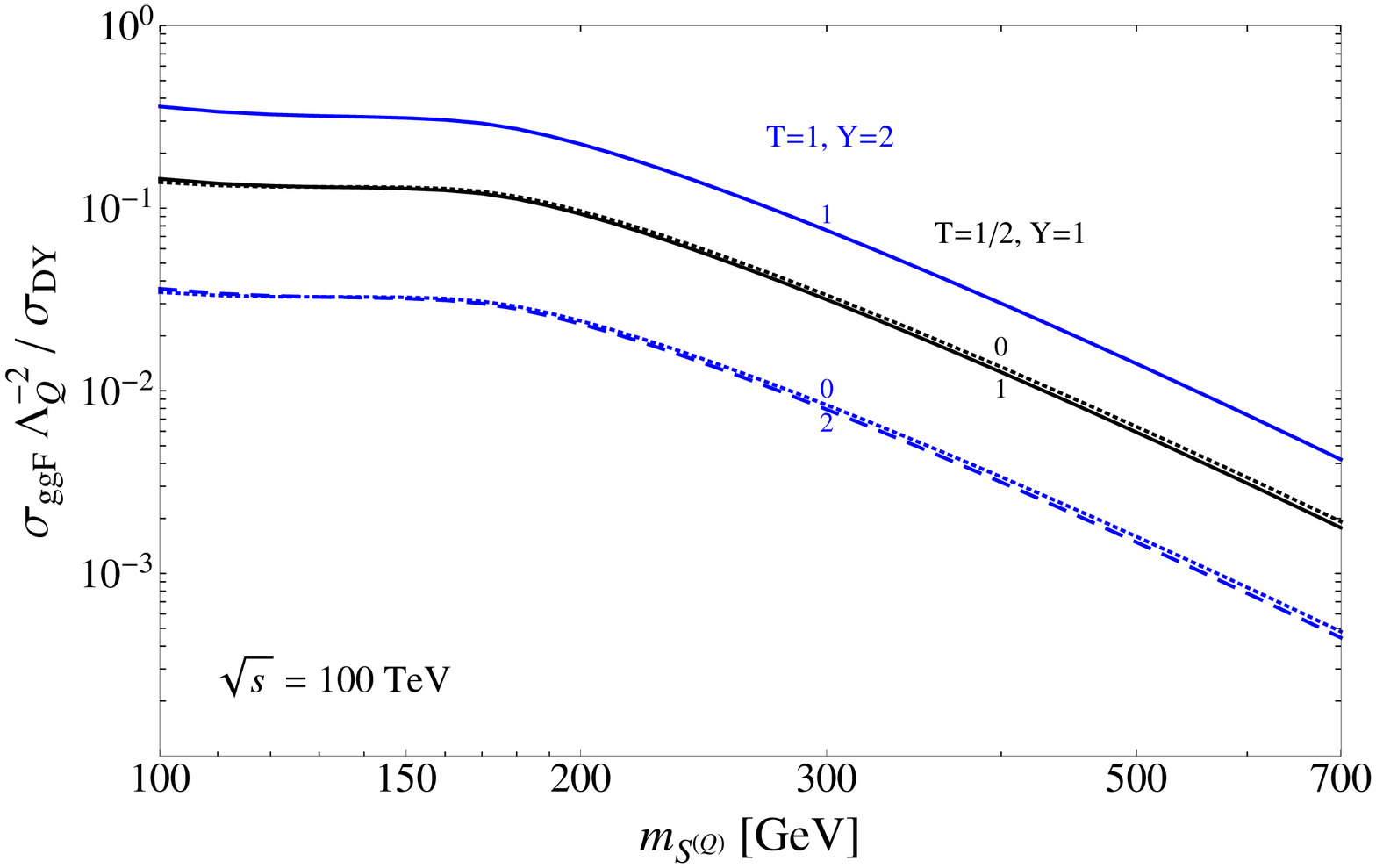} 
\includegraphics[width=0.48\textwidth]{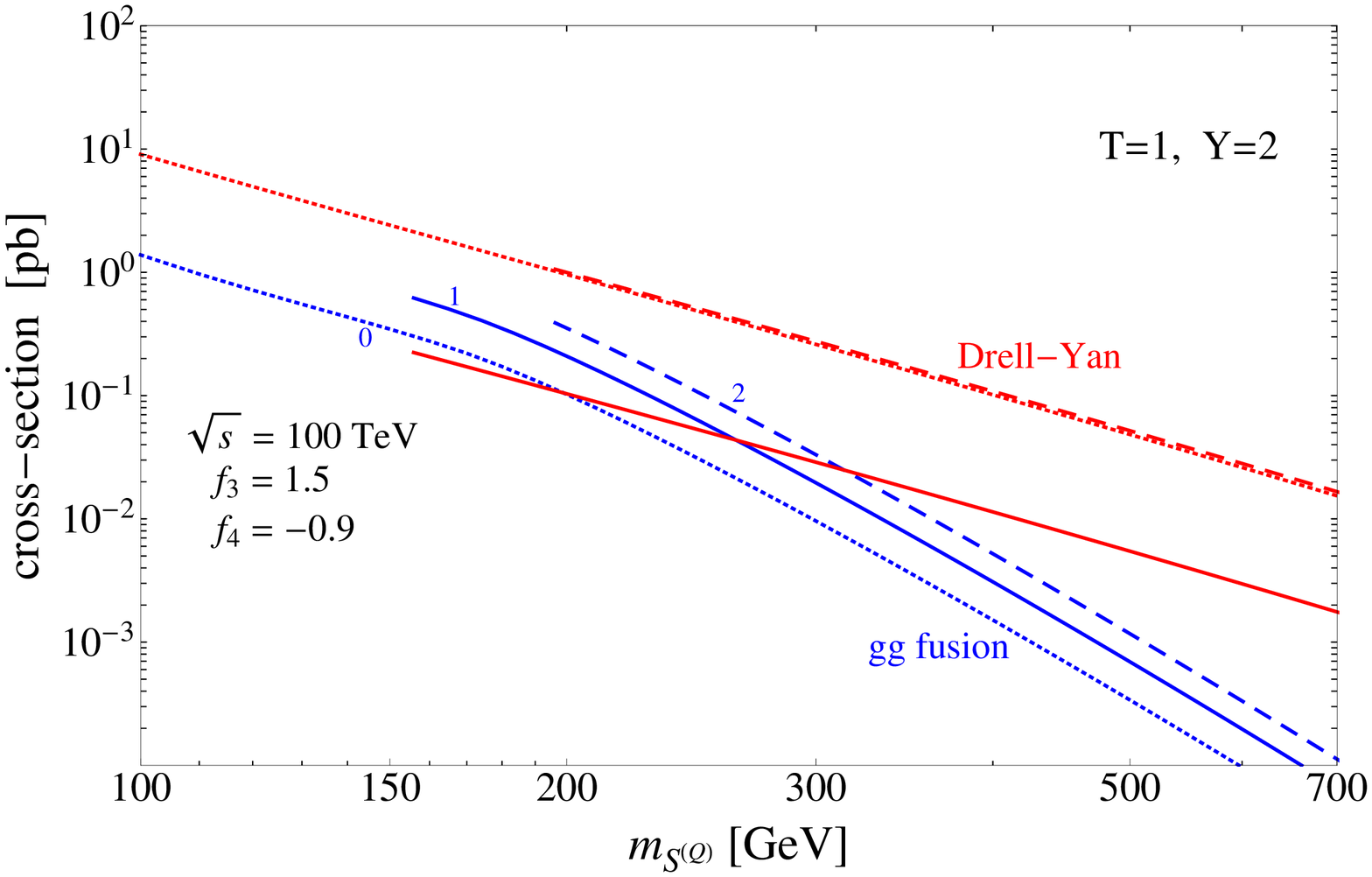}\\ 
\end{tabular}
\caption{\small{{\it Left panel:} Ratio between Higgs-mediated and Drell-Yan production cross-sections of a scalar triplet  at $\sqrt{s}=100 \TeV$. {\it Right panel:} Scalar triplet production cross-sections with  parameters as in Fig.~\ref{Fig:2} for $\sqrt{s}=100 \TeV $}}
\label{Fig:scalar100TeV}
\end{center}
\end{figure}

Qualitatively, the impact on the scalar production is similar, as we find an increase of the gluon-gluon fusion cross-section $\sigma_{\rm ggF}(100 \TeV )$  by two orders of magnitude compared to $\sigma_{\rm ggF}(14 \TeV )$ and a sizeable increase with respect to the DY process. As can be seen in Fig.~\ref{Fig:scalar100TeV},  
the larger gluon luminosity can render the gluon-gluon fusion mechanism dominant for the singly charged component of a scalar triplet, with couplings equal to the ones in Fig.~\ref{Fig:2}, up to scalar masses of 260 GeV. Furthermore, one expects exotic scalars to be copiously produced at a 100 TeV collider with a luminosity of 10 ab$^{-1}$. More specifically, we estimate that at least $10^7$ to $10^4$  pairs of singly charged scalars can~be produced for  masses  in the range $100 $ to $ 700$ GeV, even if $\Lambda_Q$ should happen to be small.

\section{Conclusions}
\label{Sec:conclusions}

We have investigated in this work the production  of exotic fermions and scalars at the Large Hadron Collider via gluon-gluon fusion, in scenarios where the extra states couple to the SM Higgs boson. More concretely, we have studied extensions of the SM by just one complex scalar field or by one fermionic field. The gauge symmetry does not restrict the gauge quantum numbers of the scalar field, while the fermionic field must be either a singlet or a $SU(2)_L$ triplet with no hypercharge and no colour, as in the type I and type III seesaw mechanisms, respectively. In many of these scenarios, then,  the extra states have also electroweak interactions and can be produced  by the Drell-Yan process with the mediation of a photon or a $Z$ boson. For the production of exotic scalar particles, we have found that the Higgs-mediated processes can give sizeable contributions to the total cross-section when the quartic couplings are $\gtrsim {\cal O}(1)$,  which can be comparable to the QCD or electroweak corrections. On the other hand, for the production of exotic fermionic particles, the relative size of both contributions does not depend on the Yukawa coupling in the minimal scenario considered in this work. We find for this case that the Higgs-mediated channel dominates the production when the mass of the fermion is $\lesssim 120$ GeV. Lastly, motivated by the current discussions on the physics opportunities at future proton-proton colliders, we have briefly addressed the prospects to produce new exotic scalars and fermions in this hypothetical machine.  We find that, in general, the Higgs interaction gives  a non-negligible contribution to the total production cross-section  and that the gluon-gluon fusion constitutes the dominant contribution to the production of heavy fermions at $\sqrt{s}= 100 \TeV$.

\section*{Acknowledgments}
The work of A. I. and E. M. is supported by the ERC Advanced Grant project ``FLAVOUR'' (267104).
This work has been partially supported by the DFG cluster of excellence ``Origin and
Structure of the Universe". S. V. also acknowledges support from the DFG Graduiertenkolleg ``Particle Physics at the Energy Frontier of New Phenomena".
A. G. H. is supported by Funda\c c\~ao para a Ci\^encia e a Tecnologia (FCT) through the grant SFRH/BD/76052/2011, financed by the European Social Fund (ESF) through POPH under the QREN framework.


\begin{thebibliography}{99}
\bibitem{CMS:2014ega}
  CMS Collaboration [CMS Collaboration],
and studies of the compatibility of its couplings with the standard model,''
  CMS-PAS-HIG-14-009.
\end{thebibliography}
\end{document}